\documentclass[runningheads]{llncs}
\usepackage{graphicx}
\usepackage{multirow}
\usepackage{todonotes}
\usepackage{comment}
\usepackage{adjustbox}
\usepackage{amsmath}
\usepackage[utf8]{inputenc}
\usepackage[english]{babel}
\usepackage{color}
\usepackage{amsfonts}
\usepackage{url}
\usepackage[mathscr]{euscript}
 \let\mathscr\relax%
\usepackage[scr]{rsfso}
\usepackage{caption}
\usepackage[labelfont=bf]{caption}

\usepackage{tabularx,booktabs}
\newcolumntype{C}{>{\centering\arraybackslash}X} 
\usepackage{multicol}
\usepackage{algorithm2e}
\usepackage{booktabs,ragged2e}
\usepackage[utf8]{inputenc}
\usepackage{todonotes}
\usepackage{lineno}
\usepackage{makecell}
\usepackage[flushleft]{threeparttable}
\usepackage{subcaption}
\usepackage{color}
\usepackage[nolist,nohyperlinks]{acronym}
\usepackage{footnote}
\usepackage[perpage]{footmisc}
\usepackage[autostyle]{csquotes}
\makesavenoteenv{tabular}
\acrodef{DL}{Deep learning}
\acrodef{CNN}{Convolutional Neural Network}
\acrodef{ML}{Machine Learning}
\acrodef{mIoU}{mean Intersection over Union}
\acrodef{GI}{gastrointestinal}
\acrodef{AI}{Artificial Intelligence} 
\acrodef{CADx}{computer aided diagnosis} 
\acrodef{CRC}{colorectal cancer}
\acrodef{DSC}{Dice Coefficient}
\acrodef{mDSC}{Dice Coefficient}
\acrodef{OOD}{Out-Of-Distribution}
\acrodef{SOTA}{State-of-the-art}
\acrodef{HD}{Hausdorff distance}
\acrodef{MDNet}{Multi decoder network}

\begin{document}

\title{MDNet: Multi-Decoder Network for  Abdominal CT Organs Segmentation}
%\title{MDNet: Multi-Decoder Network for  Liver and Spleen Segmentation}
\titlerunning{MDNet: Multi-Decoder Network for  Abdominal CT Organs Segmentation}
\author{Debesh Jha, Nikhil Kumar Tomar, Koushik Biswas, Gorkem Durak, Matthew Antalek, Zheyuan Zhang, Bin Wang, Md Mostafijur Rahman,  Hongyi Pan, Alpay Medetalibeyoglu, Vandan Gorade, Abhijit Das,  Yury Velichko, Daniela Ladner, Amir Borhani, Ulas Bagci}
\institute{Machine \& Hybrid Intelligence Lab, Department of Radiology, Northwestern University, Chicago, USA,\\ 
 Department of ECE, The University of Texas at Austin, USA}
\maketitle      
%-------------------
\begin{abstract}
%------------------
Accurate segmentation of organs from abdominal CT scans is essential for clinical applications such as diagnosis, treatment planning, and patient monitoring. To handle challenges of heterogeneity in organ shapes, sizes, and complex anatomical relationships, we propose a \textbf{\textit{\ac{MDNet}}}, an encoder-decoder network that uses the pre-trained \textit{MiT-B2} as the encoder and multiple different decoder networks. Each decoder network is connected to a different part of the encoder via a multi-scale feature enhancement dilated block. With each decoder, we increase the depth of the network iteratively and refine segmentation masks, enriching feature maps by integrating previous decoders' feature maps. To refine the feature map further, we also utilize the predicted masks from the previous decoder to the current decoder to provide spatial attention across foreground and background regions. MDNet effectively refines the segmentation mask with a high dice similarity coefficient (DSC) of 0.9013 and 0.9169 on the Liver Tumor segmentation (LiTS) and MSD Spleen datasets. Additionally, it reduces  Hausdorff distance (HD) to 3.79 for the LiTS dataset and 2.26 for the spleen segmentation dataset, underscoring the precision of MDNet in capturing the complex contours. Moreover, \textit{\ac{MDNet}} is more interpretable and robust compared to the other baseline models. The code for our architecture is available at https://github.com/xxxxx/MDNet.

 %\and Transformer
\end{abstract}

\section{Introduction}
The liver and spleen are vital organs for maintaining metabolism, immunity, and hematopoiesis~\cite{tarantino2013liver}. The liver is mainly responsible for metabolic activities, whereas the spleen controls blood filtration~\cite{wu2020generation}.  The size of the liver and spleen varies among the different populations due to age, body size, ethnicity, etc.~\cite {waelti2021normal}. Liver cancer is considered the third most common cause of cancer-related death worldwide~\cite{arnold2020global}. The spleen is one of the areas of widespread involvement of lymphoma. Computed Tomography (CT) is critical in diagnosing and characterizing liver and spleen lesions~\cite{gotra2017liver}. However, liver and spleen segmentation are among the most challenging organs to segment because of their highly variable shape, variations in morphology, location, orientation, and intensity values~\cite{gotra2017liver}.  Our clinical motivation in this work comes from the fact that the liver and spleen are part of a network that also includes the portal vein and its branches, which carry blood from the gastrointestinal tract and spleen to the liver. Therefore, there is a desire to measure liver and spleen volume, necessitating an efficient automated segmentation approach. 

% Alternate clinical motivation: The size and morphology of the liver and spleen can vary significantly based on patient's body habitus, age, and disease state. For example, portal hypertension as a result of cirrhosis may result in substantial changes in the size, contour and shape of the liver, and spelnomegaly is often the result of increased venous pressure in the splenic outflow. Other diseases, including malignancies and infections may also result in abnormal size and imaging appearance of these organs. https://pubs.rsna.org/doi/full/10.1148/rg.220025, https://pubs.rsna.org/radiographics/doi/10.1148/rg.275075023, https://pubs.rsna.org/doi/pdf/10.1148/rg.210071

There have been several studies on liver and spleen segmentation~\cite{jha2024ct,zheng2019semi,lee2022fully,demir2022transformer}. Jha et al.~\cite{jha2024ct} proposed PVTFormer architecture that used a pretrained pyramid vision transformer as an encoder and combined it with advanced residual upsampling and decoder block. They obtained a high DSC of 0.8678 and mIoU of 0.7846 for healthy liver segmentation. Lee et al.~\cite{lee2022fully} developed two deep-learning models to measure liver segmental volume ratio (LSVR) and spleen volumes from CT scans to predict cirrhosis and advanced fibrosis. In two datasets, the best-performing multivariable model achieved AUCs of 0.94  and 0.79  for cirrhosis and 0.80  and 0.71  for advanced fibrosis. They showed that the CT-based model performed similarly well to that of radiologists. Again, Demir et al.~\cite{demir2022transformer} utilized a hybrid approach by combining the Transformer with the Generative Adversarial Network (GAN) and obtained a high DSC of 0.9433.

While these architectures offer high segmentation accuracy, they might have difficulty handling variable morphology and complex anatomical structures across different individuals in the liver and spleen. Additionally, these models are complex and challenging to interpret, which limits clinical trust. This work addresses two fundamental research questions: (I) Can we develop a fully automatic, reliable, and accurate CT liver and spleen segmentation architecture? (II) Can we develop an interpretable model to visualize the model's decision-making process through the different decoder stages to enhance trust and understanding? To address the above research questions, we design a novel architecture, \acf{MDNet}. The key contribution of this work is as follows: 

\begin{enumerate}
\item \textbf{Multi-decoder learning ---} \ac{MDNet} explores the use of increasing depth in different decoders to achieve a better understanding of the liver and spleen morphology and shapes. This design allows for incremental refinement of segmentation masks, leveraging the output of the preceding decoders as spatial attention for the subsequent one, which increases the network's ability for better localization of organs.

\item  \textbf{Interpretability through visualization ---} Our architecture is designed in a way that both intermediate predicted masks and final predicted mask can be visualized through different layers. The visualization of intermediate and final masks shows the effect of different decoders, demonstrating and enabling better interpretability of the \ac{MDNet} (see -- Figure~\ref{fig:mdnet} decoder outputs and Figure~\ref{fig:qualitativeresults} (heatmap)).  
    
\item \textbf{Systematic evaluation ---} We evaluated \ac{MDNet} on two publicly available liver and spleen datasets, outperforming six SOTA segmentation methods.

%We have released all the training datasets, models, and results as open source to advance the development toward automatic polyp segmentation. 

\end{enumerate}

%Wang et al.~\cite{wang2023transliver}

\section{Method}
{\label{section:method}}
\begin{figure*} [!t]
    \centering
    \includegraphics[width=0.8\textwidth]{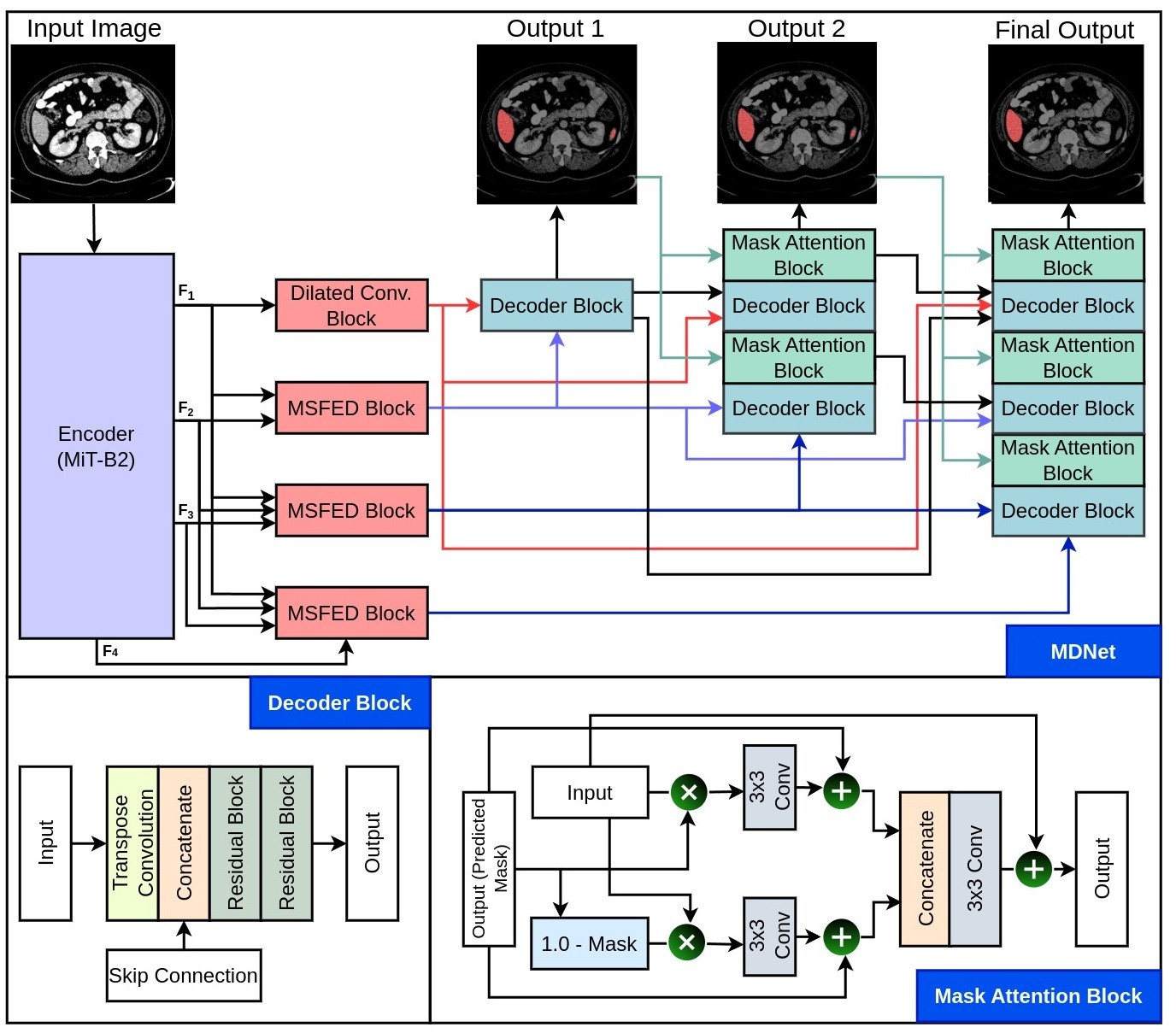}
    \caption{The block diagram shows an overview of the proposed MDNet. A MiT-B2 encoder processes the input image to extract feature maps at four different levels (F1, F2, F3, and F4). Each decoder network is connected to a different part of the decoder via a multi-scale feature enhancement dilated block to increase the depth of the network to predict three distinct segmentation masks. Additionally, the decoders are connected in a way that the output feature from the preceding decoders is utilized in the subsequent one to refine the segmentation output further. Moreover, we also use the predicted masks of the prior decoder in the subsequent decoder for the further refinement of the feature map. This process provides spatial attention across foreground and background regions and enhances the final segmentation results.}
    \label{fig:mdnet}
\end{figure*}

Figure~\ref{fig:mdnet} shows the proposed MDNet architecture schematic diagram. We employ MiT-B2~\cite{xie2021segformer} as the pre-trained encoder, from which we extract four distinct feature maps. The core component of MDNet lies in its three individual decoders, each increasing in complexity, thus progressively enhancing the depth and representational capability of the network. Additionally, MDNet integrates  Multi-Scale Feature Enhancement Dilated Blocks, Mask attention blocks, and decoder blocks to capture a richer feature representation and improve organ delineation. 

\subsection{Mix Transformer (MiT) Encoder}
Ideally, we should choose a network with high performance and balanced computational efficiency for accurate abdominal organ segmentation. Therefore, in the MDNet, we utilize MiT-B2, a variant of mix transformer, as a pretrained encoder that was previously trained in the ImageNet-1k dataset. An input image $I \in \mathbb{R}^{H\times W\times 3}$ is fed to the MiT-encoder to obtain four distinct multi-scale feature maps $F_i \in \mathbb{R}^{\frac{H}{2^{i+1}} \times \frac{W}{2^{i+1}} \times C_i}$, where  $i \in \{1, 2, 3, 4\}$ and $C_i \in \{64, 128, 320, 512\}$. These feature maps are important in capturing intricate details and varying contexts from the input abdominal images. 

\subsection{Dilated Convolution (DC) Block}
We have utilized the dilated convolution (DC) block in our architecture to enhance the feature representation of the different multi-scale feature maps from the MiT-B2 encoder. The block begins with four parallel $3 \times 3$ convolutional layers having a dilation rate of \{$1, 3, 6, 9$\}, followed by a batch normalization layer and a ReLU activation function. These dilation rates help determine the spacing between the elements in the convolutional filters, effectively enlarging the receptive field. Subsequently, the outputs from these dilated convolutional layers are concatenated to combine the multi-scale information learned at different dilation rates. This concatenated feature map is then passed through a $1 \times 1$ convolutional layer to integrate and refine the information from the multiple scales. Finally, the output from this convolutional operation is further refined using the Channel Attention and Spatial Attention mechanisms of the Convolutional Block Attention Module (CBAM)~\cite{woo2018cbam}, enhancing the network's ability to focus on relevant features while suppressing irrelevant ones, thus improving segmentation outcomes.

\subsection{Multi-Scale Feature Enhancement Dilated (MSFED) Block}
The MSFED block is a pivotal component designed to enhance feature representation by incorporating multi-scale information extracted from the MiT-encoder, which is further passed through the dilated convolution to further refine the feature representation process. With four encoder features available, the first feature directly undergoes processing through the dilated convolution block. Subsequently, the remaining three encoder feature maps are channeled into three distinct MSFED blocks, each responsible for handling a specific scale. Each MSFED block operates by taking the current scale encoder feature along with the previously processed scale features as inputs. 

% This strategic approach facilitates the extraction of multi-scale information, ensuring that the network effectively captures and leverages features across different scales for improved performance in complex image segmentation tasks.

The MSFED block for the second encoder feature map, denoted as $F_2$, is structured as follows. Initially, the feature map $F_1$ undergoes a series of $3\times3$ Conv-BN-ReLU layers, followed by a $2\times2$ max-pooling operation to reduce its spatial dimension. The resulting output from the max-pooling operation is then concatenated with the feature map $F_2$, creating a combined feature representation. Subsequently, this combined representation traverses through a series of $3\times3$ Conv-BN-ReLU layers. This sequential processing culminates in generating a unified feature representation, traversing through the dilated convolution block for further refinement. Following this initial MSFED block for the second encoder feature map, the subsequent MSFED blocks are executed in a similar manner but with more layers to accommodate more feature maps from the MiT encoder.

\subsection{Mask Attention (MA) Block}
In our MDNet architecture, the decoder component is equipped with Mask Attention (MA) blocks to facilitate spatial attention across both foreground and background regions within the feature maps. This block operates by receiving the feature maps along with the mask predicted by the preceding decoder. Initially, the input mask undergoes a resizing process followed by a sigmoid activation function, resulting in pixel values confined within the range of [0, 1]. Subsequently, two distinct masks are generated: the foreground mask, denoted as $M_f$, and the complementary background mask, represented as $M_b = 1 - M_f$. An element-wise multiplication is then performed between both the foreground and background masks and the input feature map, followed by sequential processing through a series of $3\times3$ Conv-BN-ReLU layers. Furthermore, an element-wise addition operation is executed between the feature map and the foreground mask $M_f$. The resulting attenuated feature maps are concatenated \{$\cup$\} and subsequently processed through additional $3\times3$ Conv-BN-ReLU layers, followed by a residual identity mapping connecting both input and output of the MA block.
\vspace{-2mm}
\begin{equation}
   \label{eq:mask_attention_equation_2}
F_{out} = ReLU(C_{3\times3}((C_{3\times3}(x \odot M_f) + M_f) \cup (C_{3\times3}(x \odot M_b) + M_f)) + C_{3\times3}(x))
\end{equation}

\subsection{Decoder Block}
The decoder block begins with a $2\times2$ transpose convolution layer, doubling the spatial dimensions of the feature map, which is then concatenated with the skip connection. This skip connection serves two crucial purposes: first, it aids in preserving features that may be lost in deeper layers of the network, and second, it facilitates gradient propagation during backpropagation. Following this, two residual blocks are employed to effectively learn semantic features essential for generating a segmentation mask.

\subsection{Multi-Decoder}
The decoder utilizes the feature maps from the MiT-encoder and is then processed by the MSFED block to generate the segmentation mask. All the three decoders utilize the feature maps \{$F_1$, $F_2$\}, \{$F_1$, $F_2$, $F_3$\} and \{$F_1$, $F_2$, $F_3$, $F_4$\} respectively. The initial decoder comprises a single decoder block, starting with $F_2$ as input feature maps and $F_1$ as the skip connection. Following this, the output undergoes bilinear upsampling with a factor of four, succeeded by $1\times1$ convolution and sigmoid activation to generate segmentation mask $M_1$. 

Subsequent decoders maintain a parallel structure and integrate the Mask Attention (MA) block to enhance feature representation and segmentation performance. Additionally, feature maps from previous decoders at the same scale are concatenated with a skip connection, which aids in feature propagation, thereby enhancing performance. The second decoder comprises two decoder blocks, each followed by a Mask Attention block. The resulting output from the second decoder is used to yield a segmentation mask $M_2$. Similarly, the third decoder comprises three decoder blocks, each incorporating a Mask Attention block. The output from the third decoder generates the final segmentation mask $M_3$.

% upsampled, followed by a $1\times1$ convolutional layer and sigmoid activation, culminating in the final segmentation mask $M_3$.

%The resulting output from the second decoder undergoes bilinear upsampling, followed by $1\times1$ convolution and sigmoid activation to yield segmentation mask $M_2$.

% It begins with a $2\times2$ transpose convolution layer, concatenated with the skip connection, and proceeded by two residual blocks. The semantic features from the residual blocks are augmented by the Mask Attention block, employing $M_1$ to apply foreground and background spatial attention. The output is then passed to the second decoder block, where feature maps from the first decoder and MiT-encoder are concatenated as the skip connection. This process is reiterated with another Mask Attention block utilizing $M_1$.

\section{Experiments and Results}
\subsection{{Datasets:}} We have utilized the Liver Tumor Segmentation Benchmark (LiTS)~\cite{bilic2023liver} benchmark dataset for liver segmentation. This dataset comprises $131$ CT scans for training and $70$ for testing, with only the training dataset publicly available. To avoid bias, we split the training dataset into independent training (91 scans), validation (20 scans), and test (20 scans) sets. % This distribution amounted to $13197$, $2552$, and $3414$ slices for training, validation, and testing, respectively. 
 During preprocessing, we extracted the liver masks containing healthy liver and tumors. Similarly, the Spleen segmentation dataset comes from the Medical Segmentation Decathlon~\cite{antonelli2022medical} study. This dataset includes $41$ CT scans for training and $20$ for testing. Of these $41$ training scans, the first five training scans were designated for testing, the following five for validation, and the remaining $31$ for training. %This arrangement led to the utilization of $746$ slices for training, $169$ for validation, and $136$ for testing. 
We resized the images to $512\times512$ pixels in-plane resolution to balance training time and model complexity. The volumetric CT scans were processed slice-by-slice to ensure compatibility with our computational resources. 
 
%The model was also trained, and performance was evaluated on a separate abdominal magnetic resonance imaging dataset (MRI). MRI exams from a total of 95 patients with hepatocellular carcinoma and varying levels of cirrhosis were collected ranging from 1/2010 to 1/2023. Liver and spleen segmentation masks were created by a radiologist. Image volumes and corresponding masks were resampled to standard $1 mm \times 1 mm \times 1$ mm spacing. The portal venous phase of contrast enhancement was used for segmentation. ${num}$ cases were used for training and ${num}$ cases were used for testing. A total of ${num}$ slices in the axial plane were used for training, ${num}$ cases were used for testing.

\subsection{Implementation details and evaluation metrics}
We use the PyTorch~\cite{paszke2019pytorch} framework for implementing and running all the experiments. The experiments were performed on NVIDIA RTX A6000 GPU. The network is configured to train with a batch size 16 and a learning rate set to 1e$^{-4}$. We train all the models for 500 epochs to fine-tune the network parameters adequately with an early stopping patience of 50. To enhance the performance of our network,  we use a combination of binary cross-entropy and dice loss and an Adam optimizer for parameter updates. Liver and spleen segmentation performance is evaluated using overlap-based metrics (dice coefficient (DSC), mean intersection over union (mIoU), recall, precision, and F2 score) and distance-based metrics such as Hausdorff distance (HD95).

\subsection{Results}
\textbf{{Results on LiTs datasets:}} For the baseline comparison, we compare MDNet with CNN-based architectures (U-Net~\cite{ronneberger2015u}, Attention UNet~\cite{oktay2018attention}, DeepLabv3+~\cite{chen2018encoder}) and recent transformer-based architecture such as TransUNet~\cite{chen2021transunet}, HiFormer~\cite{heidari2023hiformer} and G-Cascade~\cite{rahman2024g}. From Table~\ref{tab:results-liver}, it is evident that \ac{MDNet} surpasses all other \ac{SOTA} methods, with the highest DSC of 0.9383, mIoU of 0.9013, recall of  0.9427, and lowest HD value of 3.79. Similarly, MDNet achieves superior scores than the recent transformer-based approaches in mDSC, mIoU, recall, and HD, indicating fewer false positives. 

\textbf{{Results on MSD Spleen datasets:}} Table~\ref{tab:results-spleen} shows the results on the MSD spleen dataset. From the experimental results, it can be observed that \ac{MDNet} again outperformed all \ac{SOTA} methods with the highest \ac{DSC} value of 0.9507, mIoU of 0.9176, recall of 0.9607, F2 value of 0.9558 and HD value of 2.26. HiFormer and G-Cascade again remained the most competitive networks, with a DSC of 0.9358 and 0.9472, respectively. However, MDNet consistently outperformed them in terms of DSC, mIoU, recall, and F2 in both datasets. GCascade obtained a slightly better HD value of 0.01, which is negligible. From the overall comparison, MDNet is the optimum choice. 

\begin{figure*} [!t]
    \centering
    \includegraphics[width=0.7\textwidth]{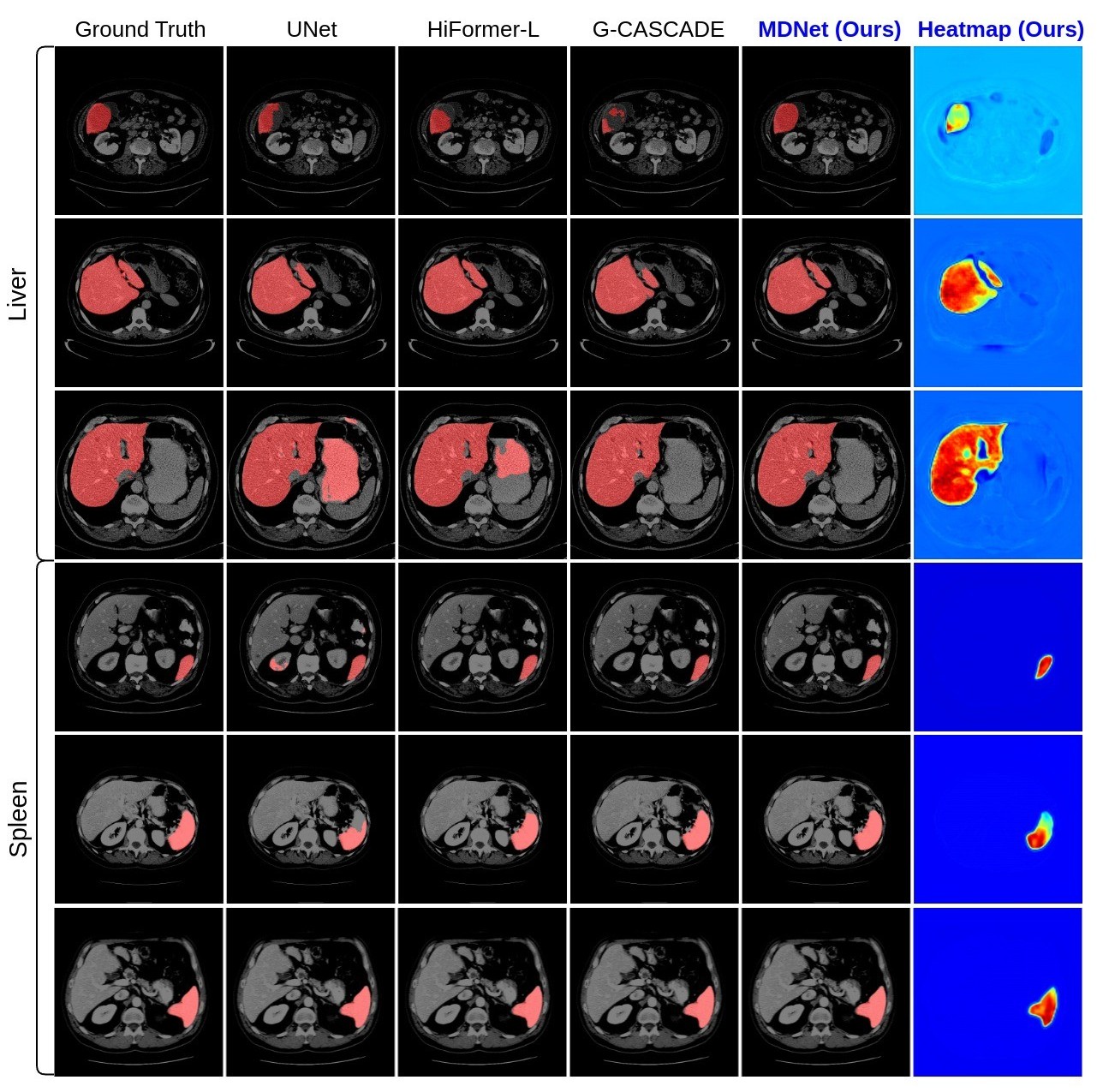}
    \caption{Qualitative comparison of models on LiTS~\cite{bilic2023liver} and MSD Spleen data~\cite{antonelli2022medical}.}
    \label{fig:qualitativeresults}
    \vspace{-3mm}
\end{figure*}

Figure~\ref{fig:qualitativeresults} shows the qualitative results. From the qualitative results, it can be seen that the segmentation output produced by MDNet closely matches the ground truth consistently as compared to UNet~\cite{ronneberger2015u}, HiFormer~\cite{heidari2023hiformer}, and G-Cascade~\cite{rahman2024g} for both liver and spleen segmentation datasets. The segmentation output produced by HiFormer and G-Cascade show over-segmentation for some cases and under-segmentation for others. Precision is important in clinical applications, and MDNet is the best choice. Table~\ref{tab:results-liver} shows the effect of MDNet on LiTS benchmark dataset. From the Table, it can be observed that there is an improvement in the performance as the network depth increases.

\begin{table}[!t]
\footnotesize
\centering
\caption{Comparative evaluation of different models on the LiTS dataset~\cite{bilic2023liver}.}
 \begin{tabular} {l|c|c|c|c|c|c|c}
\toprule
\textbf{Method} &\textbf{Publication} &\textbf{mDSC} & \textbf{mIoU}   &\textbf{Recall}& \textbf{Precision}   &\textbf{F2} &\textbf{HD}\\ 
\hline
U-Net~\cite{ronneberger2015u} &MICCAI 2015  &0.8719  &0.8153 &0.9082 &0.8724 &0.8874 &4.32 \\
Attention UNet~\cite{oktay2018attention} &MIDL 2018  &0.8975 &0.8442 &0.9015 &0.9291 &0.8975 &4.42 \\
DeepLabv3+~\cite{chen2018encoder} &ECCV 2018  &0.9295 &0.8895 &0.9214 &\textbf{0.9610} &0.9233 &3.91  \\
TransUNet~\cite{chen2021transunet} &Arxiv 2021  &0.8671  &0.8057 &0.8789 &0.8949 &0.8721 &4.69 \\
HiFormer~\cite{heidari2023hiformer} &WACV 2023  &0.9306  &0.8924 &0.9371 &0.9460 &0.9323 &3.94\\
G-Cascade~\cite{rahman2024g} &WACV 2024  &0.9301  &0.8921 &0.9287 &0.9549 &0.9271 &3.88\\
\textbf{MDNet (Ours)} &-  &\textbf{0.9383}  &\textbf{0.9013} &\textbf{0.9427} &0.9514 &\textbf{0.9380} &\textbf{3.79} \\

\bottomrule
\end{tabular}
\label{tab:results-liver}
\end{table}

\begin{table}[!t]
\footnotesize
\centering
\caption{Comparative results of different models on the MSD spleen dataset~\cite{antonelli2022medical}.}
 \begin{tabular} {l|c|c|c|c|c|c|c}
\toprule
\textbf{Method}  &\textbf{Publication}  & \textbf{mDSC}  & \textbf{mIoU}  &\textbf{Recall}& \textbf{Precision} &\textbf{F2} &\textbf{HD}\\ 
\hline
UNet~\cite{ronneberger2015u} &MICCAI 2015   &0.9126 & 0.8736 &0.9226 &0.9352 &0.9161 &2.52 \\
Attention UNet~\cite{oktay2018attention} &MIDL 2018  &0.8605  &0.8009 &0.8726 &0.8869 &0.8644 &2.90 \\
DeepLabv3+~\cite{chen2018encoder}& ECCV 2018  &0.9177 &0.8753 &0.9498 &0.9217 &0.9359 &2.41 \\
TransUNet~\cite{chen2021transunet} &Arxiv 2021   &0.8402 &0.7626 &0.8669 &0.8572 &0.8482 &3.48 \\
HiFormer-L~\cite{heidari2023hiformer} &WACV 2023  &0.9358  &0.8959 &0.9429 &0.9417 &0.9391 &2.46  \\
G-Cascade~\cite{rahman2024g}  &WACV 2024  &0.9472 &0.9135 &0.9527 &\textbf{0.9573} &0.9495 &\textbf{2.25} \\
\textbf{MDNet (Ours)} & -  &\textbf{0.9507} &\textbf{0.9176} &\textbf{0.9607} &0.9536 &\textbf{0.9558} &2.26\\
\bottomrule
\end{tabular}
\label{tab:results-spleen}
\end{table}
%-----------------

For interpretablity, we also present the heatmap in Figure~\ref{fig:qualitativeresults}. The heatmap on the example liver and spleen images shows the MDNet's precision in accurately delineating the liver and spleen boundaries. MDNet utilizes 72.33 million parameters, consumes 116.64 GMac flops, and can process at 39.70 frames per second. A detailed comparative analysis of MDNet can be found in Supp. Table 1. More visualization of MDNet output can be found in the Suppl. material .Gif file. MDNet's high performance is due to its architectural design, which integrates an iterative refinement mechanism at each decoder block, providing continuous improvement. Incorporating multi-scale feature enhancement allows MDNet to handle organ size and shape variations. The spatial attention through mask attention blocks provides constant improvement and enables the model to learn detailed and context-aware segmentation important for boundary delineation.

\begin{table}[!t]
\footnotesize
\centering
\caption{Effect of different decoders of MDNet on the LiTS dataset~\cite{bilic2023liver}.} %~\cite{bilic2023liver}
 \begin{tabular} {l|c|c|c|c|c|c}
\toprule
\textbf{Method} &\textbf{mDSC} & \textbf{mIoU}   &\textbf{Recall}& \textbf{Precision}   &\textbf{F2} &\textbf{HD}\\ 
\hline
Decoder 1 &0.9310 &0.8922 &0.9348 &0.9467 &0.9309 &3.86 \\
Decoder 2 &0.9371 &0.8991 &0.9396 &\textbf{0.9515} &0.9357 &3.80 \\
Decoder 3 &\textbf{0.9383} &\textbf{0.9013} &\textbf{0.9427} &0.9514 &\textbf{0.9380} &\textbf{3.79} \\
\bottomrule
\end{tabular}
\label{tab:results-liver}
\end{table}

\section{Conclusion}
We proposed a novel method for abdominal organ segmentation, \ac{MDNet}. Integrating a pre-trained encoder, MiT-B2, dilated convolution block, multi-scale feature enhancement dilated block, masked attention block, and decoder block, MDNet provides a robust segmentation output. Qualitative results on the two publicly available Liver CT and Spleen datasets showed that MDNet surpasses all other established baselines for liver and spleen segmentation tasks. The qualitative results and heatmap obtained by MDNet also highlighted its clinical importance. In the future, we plan to test our algorithm on MRI liver and spleen datasets built at our center. Also, volumetric computation of medical scans with MDNet and other Transformer models will be explored in the extended study.

%Our architecture can be implemented in other radiology and natural image segmentation tasks. 

\bibliographystyle{splncs04}
\bibliography{ref}
\end{document}

\title{MDNet: Multi-Decoder Network for  Abdominal CT Organs Segmentation}
%\title{MDNet: Multi-Decoder Network for  Liver and Spleen Segmentation}
\titlerunning{MDNet: Multi-Decoder Network for  Abdominal CT Organs Segmentation}

  \maketitle      
%-----------------

\begin{table*} [!htbp]
\centering
\caption{Comparative analysis of the model parameters, flops, and speed (fps) for both SOTA methods and proposed MDNet.}
\begin{tabular}{l|c|c|c|c|c} 
\toprule
\textbf{Method} &\shortstack{\textbf{Publication}\\ \textbf{Venue}} &\textbf{Backbone} & \shortstack{\textbf{Param.}\\ (\textbf{Million)}} &\shortstack{\textbf{Flops}\\\textbf{(GMac)}} & \textbf{FPS}\\ \midrule 

U-Net &MICCAI 2015 &-  &31.04	&54.75 &77.06 \\
Attention UNet &MIDM 2018 &- &8.14 &183.28 &78.81\\ 
DeepLabv3+ &ECCV 2018 &ResNet50 &39.76 &43.31 &67.65\\
TransUNet &Arxiv 2021 &CNN + ViT &67.87 &129.98 &61.38\\ 
HiFormer-L &WACV 2023 &ResNet34 + Swin Transformer &31.11 &62.09 &50.43\\
G-CASCADE &WACV 2024 &PVTv2-B2 &26.63 &22.2 &50.74\\
MDNet (Ours) &- &MiT-B2 &72.33 &116.64 &39.70\\

\bottomrule
\end{tabular}
\label{algorithm_complexity}
\end{table*}

Additionally, we provide two GIF files to visualize the output performance on Liver and spleen segmentation datasets respectively.

% \begin{table}[!htbp]
% \footnotesize
% \centering
% \caption{Comparative evaluation of MDNet's different decoders on the MSD spleen dataset.} %~\cite{antonelli2022medical}
%  \begin{tabular} {l|c|c|c|c|c|c}
% \toprule
% \textbf{Method} &\textbf{mDSC} & \textbf{mIoU} &\textbf{Recall}& \textbf{Precision}   &\textbf{F2} &\textbf{HD}\\ 
% \hline
% Decoder 1 &0.9437 &0.9053 &0.9530 &0.9478 &0.9475 &2.37 \\
% Decoder 2 &\textbf{0.9516} &\textbf{0.9176} &\textbf{0.9623} &0.9515 &\textbf{0.9568} &\textbf{2.25} \\
% Decoder 3 &0.9507 &\textbf{0.9176} &0.9607 &\textbf{0.9536} &0.9558 &2.26 \\

% \bottomrule
% \end{tabular}
% \label{tab:results-liver}
% \end{table}